\documentclass[11pt,x11names,a4paper]{article}

\usepackage[utf8]{inputenc}
\usepackage{lmodern}
\usepackage[T1]{fontenc} 
\usepackage{microtype} 

\usepackage[a4paper, left=25mm, right=25mm, top=30mm, bottom=25mm]{geometry} 

\usepackage{abstract}

\usepackage{xcolor}

\usepackage{cite}
\usepackage{hyperref}
\hypersetup{
	colorlinks=true,
	linkcolor=Blue4,
	citecolor=Red4,
	urlcolor=Green4,
	linktoc=page
}

\usepackage{amsmath,amssymb,slashed,mathbbol,mathtools}

\newcommand{\rmi}{i}
\newcommand{\rme}{\mathrm{e}}
\newcommand{\rmd}{\mathrm{d}}
\newcommand{\dd}{\mathrm{d}}

\newcommand{\be}{\begin{equation}}
\newcommand{\ee}{\end{equation}}
\newcommand{\f}{\frac}
\newcommand{\e}{\rme}

\newcommand{\C}{\mathbf{C}}
\newcommand{\Z}{\mathbf{Z}}

\newcommand{\U}{\text{U}}

\newcommand{\CP}{{\C}P}
\newcommand{\vol}{\text{vol}}

\title{\fontsize{20pt}{24pt}\selectfont\textbf{Ensembles in M-theory and Holography}\vspace{2mm}}

\author{
\large{\href{mailto:ffg1m24@soton.ac.uk}{Fri{\dh}rik Freyr Gautason}$^1$ and \href{mailto:jvanmuid@ic.ac.uk}{Jesse van Muiden}$^{2}$}\\[5mm]
{}$^1${STAG Research Centre $\&$ Mathematical Sciences, University of Southampton,}\\ 
{\normalsize University Road, Southampton SO17 1BJ, UK}\\[5mm]
{}$^2${Abdus Salam Centre for Theoretical Physics, Imperial College London}\\
{\normalsize Prince Consort Road, London SW7 2AZ, UK}\\[5mm]
}

\date{}

\begin{document}
{\hypersetup{urlcolor=black}\maketitle}
\thispagestyle{empty}

\begin{abstract}
\noindent 
We discuss that the string/M-theory partition function requires a choice of ensembles, depending on which background fields are held fixed. The background fields correspond to worldvolume couplings in the effective action approach to the superstring, which we extrapolate to the M2-brane. One natural ensemble in this context, which we call the M2-ensemble, corresponds to fixing the value of the M-theory three-form potential. In holographic setups the choice of ensemble is important when comparing to observables in the dual field theory. Indeed, in AdS$_4$ holography the M2-ensemble does not map gravitational observables directly to field theory observables at a fixed rank $N$, but rather to observables in the grand canonical ensemble. We remark that many M2-brane partition functions take a simple form in this ensemble hinting at one-loop exactness. We also discuss how in AdS$_7$ holography, the M2-ensemble does correspond to the canonical ensemble in the field theory, i.e. the (2,0) theory at fixed rank $N$. 
\end{abstract}

\maketitle

\section{Introduction and motivation}

In the mid 1980s Fradkin and Tseytlin \cite{Fradkin:1984pq,Fradkin:1985fq,Fradkin:1985ys} proposed a method to determine the generating functional of string scattering amplitudes via the path integral and a genus sum of a fundamental string in an arbitrary background specified by a metric, dilaton, and gauge potentials. In this formulation, the target space fields are viewed as sources to which the worldsheet sigma model couples and which can be used to take derivatives to obtain observables. The result is an object $\mathcal Z_\text{string}[g,B,\Phi,C]$ that can be used to generate not only the beta functions of the string (which must vanish on-shell) but also scattering amplitudes.

This generating functional is notoriously difficult to evaluate in full generality and Fradkin and Tseytlin resorted to using a derivative expansion around a genus 0 pointlike string. Focusing on the critical type II superstring, the pointlike string has vanishing on-shell action and ten bosonic zero-modes corresponding to the center-of-mass location of the string in target space. Since translation invariance is broken by the sources, the zero-mode integral is non-trivial. Furthermore, the generating functional has to be carefully regularized to obtain a sensible result \cite{Tseytlin:1988tv,Tseytlin:1988rr,Tseytlin:2006ak}. Assuming that this procedure is well-defined, target-space ${\cal N}=2$ supersymmetry fixes the form of the answer to leading order to be proportional to one of the two ten-dimensional type II supergravity actions
\be
\mathcal Z_\text{string}[g,B,\Phi,C] = - S_\text{sugra}[g,B,\Phi,C] + \cdots\,.
\ee
Here the ellipses denote perturbative corrections in $\ell_s$ and $g_s$ which so far have not been determined. Beyond perturbative corrections to this formula, we also expect non-perturbative corrections arising from string saddles with finite classical action. Including these `worldsheet instanton' corrections, the string partition function can be written as \cite{Gautason:2023igo}
\begin{equation}
\mathcal Z_{\text{string}} \approx -S_{\text{sugra}} + \sum\limits_{\text{instantons}} \rme^{-S_{\text{cl}}} Z_\text{1-loop} \,,
\end{equation}
where we have truncated all expansions at one loop (or two derivatives).

In holography, these results gain new life as its fundamental principle identifies the string generating functional to the QFT generating functional. Since, as we have summarized, the generating functional of connected string scattering amplitudes is the partition function itself, the holographic duality involves a $\log$:
\begin{equation}
\log Z_{\text{QFT}} = \mathcal Z_{\text{string}}\,.
\end{equation}
This equation is meaningless without supplementing it with a dictionary that translates between sources on the QFT side and the string theory side. Another important point is that the right hand side is not well-defined in holographic backgrounds without being regularized by including suitable boundary terms. No string theory path-integral derivation of the boundary terms exists, but instead the robust method of holographic renormalization is used to regulate and renormalize the right-hand-side (see e.g. \cite{Skenderis:2002wp}).

Without a first principle derivation, we will assume that a similar definition can be made for the M2-brane generating functional $\mathcal Z_{\text{M2}}$ by uplifting the corresponding type IIA generating functional. The former is consequently sourced by the background metric and three-form gauge potential $A_3$. Although this assumption of uplifting the type IIA observable seems ad hoc, recently there have been several non-trivial checks supporting it \cite{Beccaria:2023ujc,Beccaria:2023sph,Gautason:2025per}. The target space is now eleven-dimensional and assuming the path-integral formulation is sensible (which is far from obvious), supersymmetry fixes the form of the leading term to be proportional to the eleven-dimensional supergravity action
\be\label{Eq: M2 brane partition function in saddles}
\mathcal Z_{\text{M2}} \approx - S_\text{11D} + \sum\limits_{\text{instantons}} \rme^{-S_{\text{cl}}} (Z_\text{1-loop} + \ldots)\,,
\ee
where $S_{\text{11D}}$ contains the known two-derivative eleven-dimensional action and its corrections. Up to  two derivatives the former takes in Euclidean signature explicitly the form
\be\label{Eq: 11d action}
S_\text{11D} = \f{-2\pi}{(2\pi\ell_p)^9}\int \Big[\star \Big(R-\f{1}{2}|G_4|^2\Big)-\frac{1}{6}A_3\wedge G_4\wedge G_4 \Big]\,
\ee
where boundary terms still have to be added by hand. A point to emphasise is that the M2-brane partition function depends explicitly on the sources $g$ and $A_3$, but importantly not $A_6$. We should therefore view the partition function as defined for fixed $A_3$ (and the metric $g$) but $A_6$ is not fixed. It is then not surprising that  the leading perturbative contribution to the partition function in \eqref{Eq: 11d action} depends on $G_4$ and not its hodge dual, $G_7$. We argue that there is a conjugate ensemble where we quantize M5-branes instead of M2-branes for which $A_6$ would be fixed and $A_3$ is unfixed. It is not clear to us how to define the M5-partition function from first principles but since M5-branes are charged with respect to $A_6$, it appears that the logical microscopic description of the conjugate ensemble is in terms of M5-branes. In general, we will refer to the ensemble where one fixes the value of $A_3$ as the M2-brane ensemble, while the conjugate ensemble will be referred to as the M5-brane ensemble.  
We will also argue that in many cases one can change ensembles and therefore deduce the M5-brane partition function knowing the M2-brane partition function.

Although the formalism described above, computing $\mathcal Z_{\text{M2}}$, is expected to hold for generic smooth backgrounds, we will highlight its implications in holography, i.e. in asymptotically AdS backgrounds. In particular, we will first focus on computing gravitational observables in asymptotically AdS$_4$ backgrounds and then move on to discuss observables in asymptotically AdS$_7$ and AdS$_5$ geometries. 

Let us be more specific by discussing M-theory on AdS$_4 \times S^7/\mathbf{Z}_k$, whose holographic dual field theory is ABJM with gauge group $\text{U}(N)_k \times \text{U}(N)_{-k}$ and CS-level $k$ \cite{Aharony:2008ug}. 
The background is specified in terms $k$ and one additional parameter $L/\ell_p$ which appears both in the metric and in the three-form potential. We conclude that the M2-brane partition function evaluated on this background only depends on these two parameters. The conjugate variable to $L/\ell_p$, which defines the dual gauge potential $A_6$, is the flux quantum number $N$. This fact can be seen by rewriting the kinetic-term of $A_3$ in the leading perturbative saddle of the partition function; defining $\mu \equiv L^3/\ell_p^3$ for convenience one finds that
	\begin{equation}
		\frac{2\pi}{(2\pi \ell_p)^9} \int \star |G_4|^2 =\frac{2\pi}{(2\pi \ell_p)^9} \int G_4 \wedge G_7 = \mu N\,.
	\end{equation}
While evaluating this integral we replaced $\star G_4$ with the field strength $G_7$ and used its standard flux quantisation through the seven-sphere, the explicit expression for the four-form (see eq. \eqref{Ads4S7sol} below),  and the renormalized volume of AdS$_4$. Our interpretation is that the M2-brane partition function in this context should not be thought of as being in the \emph{ensemble} for fixed $N$ but rather for fixed $\mu$.\footnote{Note that $\mu$ is proportional to the effective dimensionless M2-brane tension, while $N$ is proportional to the effective dimensionless M5-brane tension.} One way we can think about this is that in the M2-ensemble $\mu$ is a source while $N$ is a vacuum expectation value (of the M2 number operator).

Using standard formulae, we can transform between the two ensembles. In cases where the four-form is specified in terms of a single parameter $\mu$ and its conjugate variable is an integer $N$, the transform between the two ensembles works as follows
\begin{equation}\label{Eq: discretised laplace}
\e^{\mathcal Z_{\text{M2}}(\mu)}  = \sum\limits_{N=0}^\infty \e^{\mathcal Z_{\text{M5}}(N)} \,\rme^{\mu N}\,.
\end{equation}
As already mentioned, it is not clear how to define the M5-brane partition function from first principles, or how it would reduce in a saddle point approximation to the eleven-dimensional supergravity action with $G_7$ as background source instead of $G_4$. The relation above can instead be thought of a formal definition of $\mathcal Z_{\text{M5}}$, at least in asymptotically AdS$_4$ backgrounds. 
	
The transformation in \eqref{Eq: discretised laplace} to go to the grand canonical ensemble is inverted through the inverse Laplace transform
\begin{equation}\label{Eq: conjugates of partition functions}
\e^{\mathcal Z_{\text{M5}}(N)} = \frac{1}{2\pi i}\int_{\mathcal C} \rmd \mu \, \rme^{\mathcal Z_\text{M2}(\mu) - \mu N}\,,
\end{equation}
where $\mathcal C$ is a particular contour in the complex plane that is consistent with eq. \eqref{Eq: discretised laplace}.  We note that in situations where $G_4$ is quantised and $G_7$ is defined in terms of a continuous parameter the sum and integral relations above should be interchanged.

Although a priori one is free to choose an ensemble in which to compute particular observables, the two choices clearly have an important and different interpretation. Indeed, in AdS$_4$ holography, in the M2-brane ensemble, we fix $\mu$ but $N$ is not a parameter. In order to make a comparison to a ABJM observables which are calculated for fixed $N$  and $k$, we must however compare to the M5 ensemble
	\begin{equation}
		\log Z_{\text{ABJM}}(N,k) = \mathcal Z_{\text{M5}}(N,k)\,.
	\end{equation}
The latter is more difficult to get a handle on and so we can instead compare the ABJM partition function with the Laplace transform of the M2-brane partition function
\begin{equation}\label{Eq:ABJMholoM2ensemble}
 Z_{\text{ABJM}}(N,k) = \frac{1}{2\pi i}\int_{\mathcal C} \rmd \mu \, \rme^{\mathcal Z_\text{M2}(\mu,k) - \mu N}\,.
\end{equation}
We note that $k$ appears here as a parameter in the bulk in both the M5 and the M2 ensemble as it is determined by the background metric. 
Another way of interpreting these formulae is that the M5 ensemble computes ABJM observables directly in the canonical ensemble, while the M2 ensemble  corresponds to evaluating ABJM observables in the grand canonical ensemble. 
Interestingly, in the case of the ABJM theory it was found that the partition function in the grand canonical ensemble maps to a free fermion system which significantly simplifies its evaluation \cite{Drukker:2010nc}. Furthermore, it was found to be closely related to topological string on $\mathbf{F}_0$. As we have argued, the grand canonical ensemble is the preferred ensemble for comparing to the M2-brane partition function. For the ABJM theory, several supersymmetric observables known in the literature take a simpler form in the grand canonical ensemble. We will argue that the observed simplicity can be explained by the fact that the M2-brane partition function is much more accessible and potentially simpler than the M5-brane partition function. As such, the observed simplicity does not necessarily imply an underlying topological string description.\footnote{Recently it was proposed that more generally asymptotically AdS$_4$ backgrounds in M-theory have a connection with the topological string \cite{Cassia:2025aus}. It would be interesting to see how this point of view is related to the Laplace transforms that we argue maps between M2-brane and M5-brane ensembles.} 

We emphasise that the natural ensemble used for a holographic field theory may in some cases align with the natural ensemble chosen by the M2-brane partition function. Examples highlighting this are asymptotically AdS$_7 \times S^4$ backgrounds dual to the six-dimensional (2,0) $A_{N-1}$ theory (and also the $D_N$ theories). Indeed, for these backgrounds, fixing the value of $A_3$ on the boundary fixes the $G_4$ flux quantum number through $S^4$ and so the M2-brane partition function is defined for fixed flux number $N$. The holographic dictionary therefore directly relates the M2-brane partition function to the dual QFT partition function in the canonical ensemble. Many related examples for which the M2 and canonical ensembles align can be considered by reducing the (2,0) theory on compact manifolds. Well known examples are provided by the four-dimensional class-$\mathcal S$ theories which are obtained by compactifying the (2,0) theory on a Riemann surface \cite{Gaiotto:2009we,Gaiotto:2009hg}, and their holographic duals are constructed by compactifying the associated AdS$_7$ background on the same Riemann surface \cite{Maldacena:2000mw,Gaiotto:2009gz}. 
	
In what follows we will go through a list of observables that have been computed on both sides of the holographic duality and discuss the importance of the choice of ensemble when the comparison is made. 
	
Before presenting the examples we would like to emphasise that the subtlety of choosing the correct ensemble is not novel in top-down holography. In particular in AdS$_3$ holography this has been an important aspect. Indeed, for AdS$_3 \times S^3 \times T^4$ backgrounds with pure NSNS fluxes; i.e. with fixed $Q_5$ and large $Q_1$ the dual field theory is given by the symmetric orbifold CFT of central charge $c= 6 Q_5 Q_1$. Interestingly, these backgrounds have a description in perturbative string theory for which, however, correlation functions are computed as a function of $Q_5$ and the string coupling $g_s$, not for a fixed integer $Q_1$. The resolution to this mismatch was consequently argued to come from a switch in ensembles \cite{Kutasov:1999xu,Kim:2015gak,Eberhardt:2020bgq,Aharony:2024fid}, similar to our discussion above, namely the perturbative string computes observables in the grand canonical ensemble of the dual CFT\footnote{The needed switch to change to the grand canonical ensemble was argued for most clearly in the case of $Q_5 = 1$, and for $Q_5>1$ it is expected also to hold although the exact needed transform is currently not fully understood.}
\begin{equation}
Z[p,J] = \sum_N p^N Z[N,J]\,,\quad \text{with} \quad p\propto g_s^{-2}\,.
\end{equation}
In gravity the choice of ensemble in this case is seen by either fixing the $\star H_3$-flux, corresponding to the canonical ensemble, or the boundary value of the gauge potential $B_2$, which corresponds to the grand canonical ensemble. Although for the examples discussed in this paper we do not have a perturbative string description, the observable $\mathcal Z_{\text{M2}}[g,A_3]$, which will be our main object of interest, can be computed in special cases due to supersymmetry. When comparing to the dual field theory, we find evidence that this object is indeed computed with fixed $A_3$ on the boundary, instead of $\star G_4$.

\section{Ensembles in AdS$_4$ holography}\label{sec:ensembles_in_ads_4_holography}
For completeness we start by recalling the eleven-dimensional background dual to ABJM theory at level $k$. The metric and three-form fields are given by
\begin{equation}\label{Ads4S7sol}
\dd s_{11}^2 = L^2(\dd s^2_\text{AdS} + 4 \dd \Omega_7^2)\,,\qquad G_4 = 3 i L^3 \vol_4\,,
\end{equation}
where $\dd s^2_\text{AdS}$ is the metric on Euclidean AdS$_4$ with unit radius, $\vol_4$ is its volume form, and $\dd \Omega_7^2$ is the metric on unit-radius $S^7/\Z_k$. The action of $\Z_k$ is such that it does not have any fixed points. 

\subsection{The sphere partition function} 
The simplest observable we would like to consider is where the boundary metric on AdS$_4$ is a round three-sphere. We would like to study the M2-brane partition function on this background.	As explained above, at leading order in large $L/\ell_p$, the M2-brane partition function is expected to reduce simply to the eleven-dimensional supergravity action \eqref{Eq: 11d action}. The on-shell value of this action in AdS$_4 \times S^7/\mathbf{Z}_k$ with an $S^3$ boundary is
\begin{equation}\label{Eq:M2leading}
\mathcal Z_{\text{M2}}(\mu,k)\approx -S_{\text{11D}} =  \frac{2 \mu^3}{3 \pi^2 k}\,,
\end{equation}
where $\mu\equiv (L/\ell_p)^3$. In the standard treatment we relate the length scale $L$ with the flux $N$ of the seven-form $G_7$ through the sphere.\footnote{In our conventions for Euclidean signature the definition of the dual form involves an $i$: $G_7 = i\star G_4$.} This gives $\mu^2=\f{Nk\pi^2}{2}$. Inserting this back into \eqref{Eq:M2leading} gives a result that does not match with the QFT prediction. The usual remedy is to instead compute the supergravity action written in terms of $G_7$ whose evaluation does indeed match the leading order sphere partition function of the ABJM theory (see e.g. \cite{Beccaria:2023hhi}). From the discussion above we can understand these two approaches as the difference in computing the M2-brane partition function vs. computing the M5-brane partition function. Eq. \eqref{Eq:M2leading} is the leading order contribution to the M2-brane partition function which should not be directly compared to the corresponding ABJM observable at fixed $N$. In order to compare to ABJM, we should first perform the Laplace transform \eqref{Eq:ABJMholoM2ensemble}.
For large $N$, the integral \eqref{Eq:ABJMholoM2ensemble} can be evaluated using a saddle point approximation. The dominant saddle is located at $\mu = \mu_*$ which is given by
\begin{equation}
N = \frac{\partial \mathcal Z_{\text{M2}}}{\partial \mu}\bigg|_{\mu=\mu_*} \,.
\end{equation}
To leading order, we find $\mu_*^2\approx \f{Nk\pi^2}{2}$. The saddle point contribution is now dominated by a Legendre transform
\begin{equation}
\mathcal Z_{\text{M5}}(N,k) \approx \mathcal Z_{\text{M2}}(\mu_*,k) - \mu_* N = -\f{N^{3/2}(2k)^{1/2}\pi}{3}\,,
\end{equation}
which indeed correctly reproduces the leading order behaviour of $\log Z_{\text{ABJM}}(N,k)$.

Instead of switching ensembles in gravity, we can also compare the M2-brane partition function directly with the grand canonical partition function of ABJM.
As mentioned, in  \cite{Drukker:2010nc} it was observed that the ABJM sphere partition function is in fact more easily computed in the grand canonical ensemble, defined as
\begin{equation}\label{Eq: Laplace trafo to ZQFT}
Z_\text{ABJM}(N,k) = \frac{1}{2\pi i}\int_\mathcal{C} \rmd \mu\, \rme^{J(\mu,k) - \mu N} \,,
\end{equation}
	where $J(\mu,k)$ is referred to as the grand canonical potential and $\mathcal C$ is the standard Airy contour in the complex plane.\footnote{See \cite{Marino:2011eh} for a discussion on how the contour is chosen.} By comparing to \eqref{Eq: conjugates of partition functions} we see that $J(\mu,k)$ should be directly compared with the M2-brane partition function 
	\begin{equation}\label{Eq: Z equals J}
		\mathcal Z_{\text{M2}}(\mu,k) = J(\mu,k)\,,
	\end{equation}
	consequently we will from now on refer to $J$ as being in the M2-ensemble.
	The potential $J$ can be  split into a simple perturbative piece and a non-perturbative piece
	\begin{equation}\label{Eq: grand potential}
	\begin{aligned}
		J^{}(\mu,k) =&\, J^{\text{p}}(\mu,k) + J^{\text{np}}(\mu,k)\,,\\
		J^{\text{p}}(\mu,k)=&\, \frac{C_k}{3}\mu^3 + B_k \mu + A_k\,,\\
		J^{\text{np}}(\mu,k) =&\, \sum_{m,l} f_{m,l}(\mu,k) \rme^{-(\frac{4m}{k} + 2l)\mu}\,.
	\end{aligned}
	\end{equation}
	We first focus on the perturbative part which is a remarkably simple cubic polynomial with coefficients \cite{Fuji:2011km,Marino:2011eh,Hanada:2012si,Hatsuda:2014vsa}
	\begin{equation}
	\begin{aligned}
		C_k =&\, \frac{2}{\pi^2 k}\,,\quad B_k = \frac{1}{3k} + \frac{k}{24}	\,,\\
		A_k =&\, \frac{2 \zeta(3)}{\pi^2 k} (1 - \frac{k^3}{16}) + \frac{k^2}{\pi^2} \int\limits_0^\infty \rmd x \frac{x \log (1- \rme^{-2x})}{\rme^{kx}-1}\,.
	\end{aligned}
	\end{equation}
	Notice here that the leading order behaviour of $J(\mu)$ matches the holographic computation \eqref{Eq:M2leading}. Below we will discuss why a cubic polynomial is the expected leading behaviour of an M2-brane partition function.

	Cubic polynomials, such as \eqref{Eq: grand potential}, result in an Airy function upon Laplace transformation \eqref{Eq: Laplace trafo to ZQFT}
	\begin{equation}\label{Eq: Airy function}
		Z^{\text{p}}(N,k) = C_k^{-1/3}\rme^{A_k} \text{Ai}\Big[C_k^{-1/3}(N - B_k)\Big]   \,,
	\end{equation}
where we have dropped the subscript for notational simplicity.
We see that the Laplace transform naturally gives rise to an effective shift in $N$ by the $B_k$. It has long been a puzzle what effect in string or M-theory results in this same shift. As we have seen this can be explained by the required change in ensembles when comparing M-theory results with field theory. We will discuss this in more detail below where we also show that such shifts occur when considering other observables and are again explained by the change in ensemble.

It is worth pausing here to discuss how the change of ensembles appears in the type IIA limit. In this limit we take $k\to\infty$ and reduce to ten dimensions over the Hopf fiber of $S^7$ which results in the IIA solution\footnote{We are ignoring the $B$-field which is non-zero for standard ABJM theory with two gauge nodes with unequal rank \cite{Aharony:2008gk,Aharony:2009fc}. Its presence does not play a role in our argument.}
\be
\dd s_{10}^2 = L_{10}^2(\dd s_\text{AdS}^2 + 4\dd s^2_{\CP^3})\,, \quad F_2 = \f{4L_{10}}{g_s}J\,,\quad F_4 = \f{3iL_{10}^3}{ g_s}\vol_4\,,
\ee
where we have written the solution in string theory language, meaning the string length $\ell_s$ and string coupling $g_s$. In the standard treatment we would now relate the ten-dimensional length scale to the effective coupling constant $\lambda=N/k$ through $L_{10}^2 = \pi\ell_s^2\sqrt{2\lambda}$ and then quantize the 6-form flux through $\CP^3$ leading to a relation between the string coupling constant $g_s$ and the flux number $N$ given by $(g_sN)^4 = \pi^2(2\lambda)^{5}$. However, just as the M2-brane partition function is not a function of $N$, the string partition function is also not. Instead it depends directly on $g_s$ and once again the IIA string description is not dual to a single ABJM theory with fixed $N$ but rather the grand canonical ensemble. In order to make connection to the field theory we want to switch ensembles, roughly speaking from $g_s$ to $N$. When doing so however, we must take into account that the flux of $F_2$, which is related to the Chern-Simons level $k$, should be held fixed
\be\label{F2flux}
k = \f{1}{2\pi \ell_s}\int_{\CP^1} F_2 = \f{2L_{10}}{\ell_s g_s}\,.
\ee
The dual variable to $N$ can be deduced from the $F_4$ kinetic term which for our solution evaluates to
\be
\f{2\pi}{(2\pi\ell_s)^8}\int \star_{10}|F_4|^2 = \f{L_{10}^3}{\ell_s^3 g_s}N\equiv \mu N\,,
\ee
where we have once again used flux quantisation to determine the relevant conjugate variable. Notice that here $\mu=L_{10}^3/(\ell_s^3 g_s)$ similarly to what we found in M-theory but with a slight difference due to the string coupling constant. To compute the leading order contribution to the type IIA string partition function we evaluate the (Euclidean) ten-dimensional supergravity action
\be
S_\text{10D} = \f{-2\pi}{(2\pi \ell_s)^8} \int\star_{10}\bigg[\e^{-2\Phi}\Big( R+ 4 |\dd\Phi|^2 -\f12 |H_3|^2\Big)-\f12|F_4|^2-\f12 |F_2|^2\bigg] + \text{CS-terms}\,,
\ee
on shell as we did in eleven dimensions. The only contribution comes from the RR-form kinetic terms which can be regularized by assigning the standard regularized volume to AdS$_4$ with a three-sphere boundary. The result is
\be
{\cal Z}_\text{string} \approx -S_\text{10D} =\f{L_{10}^8}{3 \pi^2 g_s^2 \ell_s^8}\,.
\ee
Replacing the string theory parameters $L_{10}$ and $g_s$ with the QFT parameters using flux quantization does not yield a match with the QFT due to the fact that we are in the wrong ensemble. As discussed above we must perform a Laplace transform. First we must rewrite the expression in terms of the fixed quantity $k$ using \eqref{F2flux}  and the quantity $\mu$. In this way, we recover exactly the result in \eqref{Eq:M2leading} and the rest of the argument follows analogously.

	\subsection{General expectations for the perturbative M2-partition function} 
	Recently a large body of literature has been dedicated to studying supersymmetric partition functions for general $\mathcal N=2$ CS-matter theories for which the CS levels add up to zero. It was conjectured that the perturbative part of these partition functions all take the form of an Airy function whose coefficients depend on the particular theory and observable \cite{Marino:2011eh,Bobev:2022eus,Hristov:2022lcw}
	%
	\begin{equation}\label{generalAiry}
		Z^{\text{p}}(N,k,\mathbf{q}) = C_k(\mathbf{q})^{-1/3}\rme^{A_k(\mathbf{q})} \text{Ai}\Big[C_k(\mathbf{q})^{1/3}(N - B_k(\mathbf{q}))\Big]   \,,
	\end{equation}
	where $\mathbf{q}$ encodes both the field theory and background parameters determining the particular observable that is being computed. A summary of the evidence in favor of this Airy conjecture can be found in \cite{Bobev:2025ltz}, and references therein. For a long time it has been a mystery how this answer should be reproduced from gravity. A natural expectation is that the M-theory computation involves infinitely many terms corresponding to the series expansion of the Airy function at large $N$. As we have argued, when computing the M2-brane partition function, we should not compare directly to \eqref{generalAiry}, but instead the corresponding grand canonical partition function. Thus transforming \eqref{generalAiry} to the grand canonical ensemble and identifying with a M2-brane partition function in the corresponding eleven-dimensional geometry, we conclude that the full perturbative part of the M2 partition function (namely excluding instantons), is given by a cubic polynomial 
	\begin{equation}\label{Eq: cubic polynomial}
		\mathcal Z_{\text{M2}}^{\text{p}}(\mu,k,\mathbf{q})=\, \frac{C_k(\mathbf{q})}{3}\mu^3 + B_k(\mathbf{q}) \mu + A_k(\mathbf{q})\,.
	\end{equation}
	This simple result implies that apart from the leading two derivative supergravity contribution to the perturbative M2-brane saddle (see \eqref{Eq: M2 brane partition function in saddles}), there are only two perturbative corrections. In principle these could be determined by computing higher loop corrections around the degenerate M2-brane saddle, but this is currently technically out of reach. These corrections are expected to give rise to higher-derivative corrections to the leading order two derivative supergravity action.\footnote{We are discussing higher-derivative corrections to eleven-dimensional supergravity. One can also study higher derivative corrections in four-dimensions (see e.g. \cite{Bobev:2021oku}) but it appears to us that these would be relevant when computing the partition function in the M5-ensemble but not the M2-ensemble.}  For the consistency of anomaly cancellations in type II string theory it is known that the eleven-dimensional action gets corrected by a topological term commonly referred to as $A_3 \wedge I_8$ \cite{Vafa:1995fj,Duff:1995wd}, which contributes at eight derivative order. A second term that also contributes at eight derivatives is commonly denoted as $t_8 t_8 R^4$ \cite{Green:1997di,Green:1997as}. The supersymmetric completion of these terms is not fully understood but it is expected that only two superinvariants exist at this derivative order \cite{Tseytlin:2000sf}. Importantly there is no correction at four or six derivative order, and so the eight derivative terms are the leading correction to the two-derivative action. Counting derivatives, these terms lead to the  $\mu^1$ contribution in \eqref{Eq: cubic polynomial} and most likely both terms play a role  (See \cite{Bergman:2009zh} for the evaluation of the anomaly term). A higher derivative correction that could explain the $\mu^0$ term is currently unknown (or even if it should be interpreted as a local higher derivative term, see e.g. the discussion in Section 3.2 of \cite{Beccaria:2023hhi}). Since it is independent of $\mu$, this term is an additive factor in both the M2 and M5 ensemble partition functions, but generically it depends on other parameters $\mathbf{q}$. In principle there is nothing that prevents further higher-derivative corrections that would give rise to $1/\mu$ correction in the M2-brane partition function. However, taking the Airy conjecture above for granted, there seems to be some persistent feature which eliminates all $1/\mu$ corrections. We do not have a M-theoretic explanation for why the perturbative series truncates at $\mu^0$ but giving such an argument is tantamount to proving the Airy conjecture. It is interesting to note that the M2-brane partition function computed in backgrounds dual to the six-dimensional (2,0) theory (and compactifications) seem to exhibit the same cubic structure, something we highlight in more detail in the sections below. We expect that the same explanation is at play in those examples as in the AdS$_4$ backgrounds studied here. 
	\subsection{Non-perturbative corrections}
	Now we turn to the non-perturbative contributions to the grand canonical potential in the large $\mu$ expansion, and for technical simplicity we mostly constrain ourselves to the sphere partition function of the ABJM theory. The non-perturbative terms come in three forms \cite{Drukker:2011zy}
	\begin{equation}
	\begin{aligned}
		\text{worldsheet instantons} &\sim  f_{m,0}(k) \rme^{-4m\mu/k}\,,\\
		\text{membrane instantons} &\sim f_{0,l}(\mu,k) \rme^{-2l\mu}\,,
	\end{aligned}
	\end{equation}
	and finally bound states that combine both exponential behaviors \cite{Hatsuda:2013gj}. The worldsheet instanton coefficients depend on $k$, but not on $\mu$, and can be written in terms of of Gopakumar-Vafa invariants on $\mathbf{F}_0$ \cite{Hatsuda:2012dt}:
	\begin{equation}\label{Eq: WS instanton prefactors}
		f_{m,0}(k) = \sum_{g \geq 0}\sum_{d p = m} \sum_{d} \frac{n_g^{d}}{p\left(2 \sin \frac{2 \pi p}{k} \right)^{2-2g}}\,.
	\end{equation}
	For future reference we would like to single out a particular tower of instanton contributions, with the labels $(g,d) = (0,1)$, for which their coefficients equal
	\begin{equation}\label{Eq: WS instanton prefactors special}
		f_{m,0}^{g=0,d=1}(k) =  \frac{n^{1}_0}{4 m \sin^2 \frac{2 \pi m}{k}}\,.
	\end{equation}
	Interestingly, we will see the same answer, modulo the Gopakumar-Vafa invariant, arise in different holographic setups in the following sections.\footnote{It is expected that the GV invariants arise from a supersymmetry constraint on the embeddings of the brane, and should thus not be taken into account when comparing instantons in different holographic backgrounds.} The coefficients for the membrane and bound state instantons have not been determined in closed form but are known to be a second degree polynomial in $\mu$
	\begin{equation}\label{Eq: coefficients membrane instantons}
		f_{m,l\neq 0}(\mu,k) = a_{m,l}(k)\mu^2 + b_{m,l}(k)\mu + c_{m,l}(k)\,,
	\end{equation}
	and have been explicitly computed for several instanton numbers \cite{Hatsuda:2013gj,Hatsuda:2013oxa}. Recently there has also been progress on the gravitational side of the duality. The leading worldsheet instanton was computed through a semi-classical quantisation of the type IIA string/M2-brane wrapping a compact cycle inside AdS$_4 \times S^7/\mathbf{Z}_k$ \cite{Gautason:2023igo,Beccaria:2023ujc}:	
	\begin{equation}\label{Eq: instanton ABJM}
		\mathcal Z^{(1,0)}_{\text{M2}}(\mu,k) = \frac{1}{\sin^2 \frac{2\pi}{k}} \rme^{-4\mu/k}\,,
	\end{equation}
	where we have expressed the answer in terms of $\mu$ and not $N$ as in the original works \cite{Gautason:2023igo,Beccaria:2023ujc}.	We now observe by comparing to the field theory result for $J^{\text{np}}(\mu,k)$ that this one-loop M2-partition function is exact, i.e. there are no $1/\mu$ corrections.  In the canonical ensemble on the other hand, fixing $N$ instead of $\mu$, one finds that the corresponding non-perturbative correction is significantly more involved and receives infinitely many $1/N$ corrections. To see this one combines the instanton partition function with its perturbative counterpart and then performs the Laplace transform. The leading non-perturbative effect at large $N$ can then be determined to be \cite{Hatsuda:2012dt}
	\begin{equation}
		\mathcal Z^{(1,0)}_{\text{M2}}(N,k) = \frac{1}{\sin^2 \frac{2\pi}{k}} \frac{Z^{\text{p}}(N+4/k,k)}{Z^{\text{p}}(N,k)} \,,
	\end{equation}
	where $Z^{\text{p}}$ was defined in \eqref{Eq: Airy function}. The reason that this result takes the form of a ratio of perturbative partition functions $Z^\text{p}$, with a shift in the argument of the numerator, is a consequence of the fact that the on-shell action of the M2-brane scales linearly in the M2-ensemble: $S_\text{cl} = \sigma_{\text{cl}}\,  \mu$ for some $\mu$-independent quantity $\sigma_\text{cl}$. Upon Laplace transformation to the canonical ensemble this results in the shift in the argument of the numerator of the form $N+\sigma_{\text{cl}}$.
Recently, we have studied non-perturbative M2-brane corrections to a large class of supersymmetric ABJM observables, generalising the answer above. These include the superconformal index, topologically twisted index, and squashed sphere partition function \cite{Gautason:2025per}. The one-loop M2-brane partition function is universally written as
	\begin{equation}\label{Eq: final answer}
	\mathcal Z^{(1,0)}_{\text{M2}}(\mu,k,x_{\pm}) = 2 \sum\limits_{\text{fixed points}} \frac{s(\tfrac{2}{k})^{2k}s(x_+)^{-k}s(x_-)^{-k}}{t(x_+)t(x_-)}\rme^{-4\mu/k}\,,
	\end{equation}
	where the sum runs over the fixed points of a background Killing vector,\footnote{This localisation to fixed points follows from the localisation of zero-modes in the M2-brane partition function, and is related to the localisation of the supergravity action that was first discussed in \cite{BenettiGenolini:2019jdz}.} $x_\pm$ are background parameters that merely specify the particular boundary partition function, and finally the functions $s$ and $t$ are standard contributions to three-dimensional supersymmetric partition functions:
	\begin{equation}
	\begin{aligned}
		s(z) =&\, \rme^{\frac{\rmi  \text{Li}_2(\rme^{2\pi\rmi z})}{2\pi}- \frac{\rmi \pi}{12}  - z\log\left[ 1-\rme^{2\pi\rmi z} \right]+\frac{\rmi\pi z^2}{2}}\,,\\
		t(z)=&\, \frac{4}{k}\frac{\sin\pi z}{z}\,.
	\end{aligned}
	\end{equation}
	We argued in \cite{Gautason:2025per} that this one-loop partition function of instantonic M2-brane saddles is one-loop exact in the M2 ensemble (the grand canonical ensemble) implying that there are no $1/\mu$ corrections. A similar one-loop exact behaviour was seen for the non-perturbative grand canonical potential from a field theory computation of the (squashed) sphere partition function of mass-deformed ABJM \cite{Hatsuda:2016uqa,Nosaka:2024gle,Kubo:2024qhq}.

	\subsection{Wilson loops}
	It was vital to understand the relationship between the field theory ensemble and bulk M2-brane ensemble to appreciate the exactness of the one-loop partition functions of the subleading M2-brane saddles in \eqref{Eq: M2 brane partition function in saddles}. Now we argue that one-loop partition functions of non-degenerate supersymmetric M2-branes are exact in more general setups, and in particular in the context of Wilson loop expectation values. Expectation values of operators are transformed between canonical and grand canonical ensembles through the usual Laplace transform:
	\begin{equation}
		\langle \mathcal O(N,k) \rangle = \frac{1}{Z(N,k)} \frac{1}{2\pi i} \int_{\mathcal C} \rmd \mu \, \rme^{J(\mu,k) - \mu N} \langle \mathcal O(\mu,k)\rangle\,.
	\end{equation}
	The study of supersymmetric Wilson line operators in ABJM was carried out in \cite{Klemm:2012ii}. In particular, the expectation value of the 1/2-BPS Wilson line in the fundamental representation was found to be\footnote{We have taken the normalisation of \cite{Giombi:2020mhz,Giombi:2023vzu}, which differs by a factor $1/2$ compared to \cite{Klemm:2012ii}.}
	\begin{equation}\label{WLM5ensemble}
		\langle W(N,k) \rangle = \frac{1}{2 \sin \frac{2\pi}{k}}\frac{Z^{\text{p}}(N-2/k,k)}{Z^{\text{p}}(N,k)} + \mathcal O(\rme^{-N})\,.
	\end{equation}
	In the M2-ensemble (i.e. the grand canonical ensemble) this answer translates to a particularly simple expression
	\begin{equation}\label{Eq: WL ABJM}
		\langle W(\mu,k) \rangle = \frac{1}{2 \sin \frac{2\pi}{k}} \rme^{2\mu/k} + \mathcal O(\rme^{-\mu})\,.
	\end{equation}
	We can compare this to a recently computed one-loop partition function of an M2-brane wrapping an AdS$_2 \times S^1/\Z_k \subset \text{AdS}_{4}\times S^{7}/\Z_k$ \cite{Giombi:2023vzu}. We note that from the perspective of keeping $\mu$ fixed, this partition function is one-loop exact. Once this is transformed to the M5-ensemble \eqref{WLM5ensemble}, it is a complicated expression with infinitely many $1/N$ corrections to the leading order behaviour. 	
	
We note that the shift in the argument of the partition function in \eqref{WLM5ensemble} is controlled by the classical M2-brane action. Assuming that the M2-brane partition function for Wilson loops in higher representations is also one-loop exact, we can predict the form of the ABJM observable up to a prefactor given by the one-loop action. For example, for the multi-wound M2-brane on AdS$_2\times S^1/\mathbf Z_k$ the classical action is given by
\begin{equation}
S_{\text{cl}} = - \f{2n\mu}{k}\,,\quad\Rightarrow\quad \langle W_{n}(\mu,k) \rangle = Z_\text{1-loop} \e^{2n\mu/k} 
\end{equation}
where $n$ is the winding number and $Z_\text{1-loop}$ is the one-loop partition function of the M2-brane which so far has not been computed. Assuming that the one-loop action is independent of $\mu$, we can  use the Laplace transform \eqref{WLM5ensemble} to argue that the Wilson loop expectation value in the canonical ensemble should take the form
\begin{equation}
\langle W_{n}(N,k) \rangle \propto\frac{Z^{\text{p}}(N-2n/k,k)}{Z^{\text{p}}(N,k)} + \mathcal O(\rme^{-N})\,,
\end{equation}
where the proportionality factor is determined by the M2-brane one-loop partition function. We can compare this to the localisation answer for a multi-wound Wilson loop obtained directly in field theory \cite{Klemm:2012ii}\footnote{This multi-wound Wilson loop is a sum of representations whose Young tableau has in total $n$ boxes; $n-s$ boxes in the first row and one box in the remaining $s$ rows: $\langle W_n \rangle = \sum_{s=0}^{n-1} \langle W_{n,s}\rangle$. }
\begin{equation}\label{Eq: multi wound WL ABJM}
\langle W_{n}(N,k) \rangle = \frac{1}{2 \sin \frac{2\pi n}{k}}\frac{Z^{\text{p}}(N-2n/k,k)}{Z^{\text{p}}(N,k)} + \mathcal O(\rme^{-N})\,,
\end{equation}
which shows a nice agreement and confirms the change of ensembles advocated for in this paper. Furthermore we find from this field theory result, similar to the multi-wound instanton discussed in the previous section \eqref{Eq: WS instanton prefactors special}, that the one-loop partition function of the multi-wound supersymmetric M2-brane gets a particularly simple modification compared to the single-wound brane:
\begin{equation}
	\mathcal Z_{\text{M2}}^{(n)}(\mu,k) =  \mathcal Z_{\text{M2}}^{(1)}(\mu,k/n)\,,
\end{equation}
where $n$ is the winding number of the M2-brane. The difference between the multi-wound instanton and the multi-wound Wilson loop, when written as a function of the single-wound brane, is a factor $n^{1-\chi}$, where $\chi$ is the Euler number of the base space over which the M-theory circle is fibred. This result is consistent with the conjecture made in \cite{Gautason:2023igo} upon reduction to type IIA string theory, and we will see in the following sections that it is consistent in other holographic setups as well. This leads us to proposing the following general relation between the one-loop partition functions of multi-wound and single-wound supersymmetric M2-branes that wrap the M-theory circle
\begin{equation}\label{Eq: multiwound vs single wound}
	\mathcal Z_{\text{M2}}^{(n)}(\mu,k) =  n^{1-\chi}\mathcal Z_{\text{M2}}^{(1)}(\mu,k/n)\,,
\end{equation}
where the parameter $k$ controls the size of the M-theory circle.

Similar expressions for Wilson loops in other representations are available in the literature \cite{Hatsuda:2016rmv} that relate their expectation values to shifted partition functions, like in \eqref{Eq: multi wound WL ABJM}. It would be very interesting to explain these results directly from the gravitational theory through a correct embedding of M2-branes, which we leave for future work.

The remarkable simplicity of M2-brane partition functions that result in the simple polynomial structure in the case of the degenerate M2-brane and the membrane instantons as well as the even simpler one-loop exact form for worldsheet instanton and Wilson loops begs for an explanation. One possibility is that supersymmetric localisation can be performed on the M2 worldvolume which explains all of these features. Approaching such an explanation requires finding a suitable off-shell fermionic charge in the worldvolume theory but as far as we are aware no suitable candidate has been identified. These examples further hint at the existence of one-loop exact M2 partition functions which have not been computed directly in gravity.

\subsection{The giant graviton expansion}
To close our discussion on AdS$_4$ holography we briefly mention the supersymmetric index of the ABJM theory\cite{Bhattacharya:2008zy,Bhattacharya:2008bja}. The simplest index we can consider is the index which counts 1/2-BPS operators which takes the form
\be
I_N(q) = \f{1}{(q;q)_N}\,,\qquad (a;q)_n = \prod_{k=0}^{n-1}(1-aq^k)\,.
\ee
Here $q = \e^{-\beta}$ where $\beta$ is the lenth of the thermal cycle. Recently this and other indices have been expressed in a way that is suggestive for its holographic description \cite{Arai:2020uwd,Gaiotto:2021xce}
\be
\f{I_N(q)}{I_\infty(q)} = 1+\sum_{n=1}^\infty \f{1}{(q^{-1};q^{-1})_n}q^{nN}\,.
\ee
The gravitational interpretation of this formula is that $I_\infty$ is the index of graviton states in AdS$_4$ whereas the sum on the right-hand-side represents the contribution of M5 brane giant gravitons \cite{McGreevy:2000cw}. In this sense the giant graviton classical action is given by $n\beta N$ and the coefficient $(q^{-1};q^{-1})_n^{-1}$ is interpreted as the worldvolume index on the M5-brane giant albeit with the fugacities reversed when compared to the original ABJM index. This has been explicitly verified by a semi-classical quantization of an M5 brane with $S^1\times S^5$ worldvolume in AdS$_4\times S^7$ \cite{Beccaria:2023cuo}.

When interpreted as an M-theory partition function, this result (and other less supersymmetric indices)  conforms with the message of this paper. The canonical partition function (or index in this case) is dual to an M5-brane partition function. On the other hand the index in the grand canonical ensemble
\be
\Xi_\mu = \sum_{N=0}^\infty \e^{\mu N} I_N(q) = \f{1}{(\e^{\mu},q)_\infty}\,, 
\ee
should be instead related to the M2-brane partition function. We expect that dual giant gravitons will play a prominent role in the gravitational expression as dual giants are indeed realized by M2-branes in AdS$_4$ \cite{Grisaru:2000zn,Hashimoto:2000zp}. We leave exploring the relation between the grand canonical index, dual giant gravitons, and the M2 partition function on AdS$_4$ for future explorations.
\section{Ensembles in AdS$_7$ holography}
In this section we discuss holographic examples where the M2-ensemble does correspond to the canonical ensemble in the dual field theory. These are asymptotically AdS$_7$ backgrounds in M-theory and other backgrounds which are related to the backreaction of M5-branes. We will start by considering the near horizon limit of M5-branes probing $\mathbf{C}^2/\Gamma \times \mathbf{R}$ singularities giving rise to the six-dimensional (2,0) SCFTs which admit an ADE classification. For simplicity we will focus only on the cases where $\Gamma$ is either trivial corresponding to the $A_{N-1}$ series or $\mathbf Z_2$ corresponding to the $D_N$ series. Due to the strongly coupled nature of the CFTs, only few tools are available for their direct study. Rare exceptions include their anomalies, supersymmetric indices, and particular defect operators which we will discuss below.

Another tool at our disposal is of course holography where the dual geometries are respectively AdS$_7 \times S^4$ for the $A_{N-1}$ series and AdS$_7 \times \mathbf{RP}^4$ for the $D_N$ series.  The M2-brane generating functional $\mathcal Z_{\text{M2}}$ is as before defined in terms of a fixed three-form potential, thus fixing the four form flux quantum number through the internal space. Consequently, the rank of the gauge group in the dual field theory is fixed and thus the M2-brane partition function corresponds to the canonical partition function in the QFT. 
	
We note that in this section we discuss well-known facts about M-theory in asymptotically AdS$_7$ space-times. Our main motivation to review these aspects is to highlight the resemblance between the results available for AdS$_7$ and AdS$_4$ backgrounds, crucially once they are both compared in the M2-ensemble.

	\subsection{Anomalies} The Weyl anomaly of six-dimensional SCFTs involves two combinations of curvature tensors and is controlled by the coefficients $a$ and $c$ as follows 
	\begin{equation}
		\mathcal A_6 = a \, E_6 + c \, W_6\,,
	\end{equation}
	where $E_6$ represents the six-dimensional Euler density and $W_6$ is a linear combination of the three independent contractions of the Weyl tensor. Even though there are three independent Weyl invariant terms, because of the large number of supersymmetry, there is only a single $c$-anomaly coefficient. For the $A_{N-1}$ theories the $a$ and $c$ coefficients were computed in \cite{Harvey:1998bx} where it was found that they take cubic forms\footnote{The $N^0$ term can be fixed by forcing the empty theory to have vanishing anomaly coefficients.}
	\begin{equation}
		a_{A_{N-1}} = 4N^3 - \frac{9}{4} N - \frac{7}{4}\,,\qquad c_{A_{N-1}} = 4 N^3 - 3N  -1\,.
	\end{equation}
	In holography these anomaly coefficients can be computed in gravity by evaluating an on-shell action as a function of the boundary metric \cite{Henningson:1998gx}. As argued in the introduction, this should be viewed as the leading order contribution to the M2-brane partition function. The fact that the anomaly coefficients are cubic polynomials in $N$ is in good agreement with this interpretation. Indeed, from the supergravity point of view the structure of this polynomial was argued to arise from a leading supergravity contribution, its eight derivative corrections $t_8 t_8 R^4$ and $A_3 \wedge I_8$, and a one-loop supergravity computation \cite{Henningson:1998gx,Tseytlin:2000sf,Bastianelli:2000hi,Beccaria:2014qea}. It is interesting to note that the anomaly coefficients are cubic polynomials and do not receive $1/N$ corrections exactly analogous to the grand canonical partition function of the ABJM theory.
	
	Interestingly, for the $D_N$ series there is an additional quadratic term \cite{Intriligator:2000eq,Yi:2001bz}. However, upon identifying the correct M-theory variable (namely, the four-form involves a shift) we once again obtain a cubic polynomial with no quadratic term:
	\begin{equation}
		a_{D_N} = 4\big(2N-1\big)^3 - \frac{9}{4} \big(2N-1\big) + \f74\,,\quad c_{D_N} = 4 \big(2N-1\big)^3 - 3 \big(2N-1\big) + 1\,.
	\end{equation}

	Another class of examples is obtained by reducing the (2,0) $A_{N-1}$ theory on a genus-$g$ Riemann surface $\Sigma_g$. The resulting infrared theories are four-dimensional ${\cal N}=2$ class-$\mathcal S$ of Gaiotto \cite{Gaiotto:2009we,Gaiotto:2009hg}. The holographic dual geometry is of the form ($\text{AdS}_5\times \Sigma_g) \times \tilde S^4$, where the four sphere is squashed and fibred over the Riemann surface \cite{Maldacena:2000mw,Gaiotto:2009gz}. The conformal anomalies of the  class-$\mathcal S$  theories can be computed by appropriately integrating the anomaly polynomial of the parent six-dimensional theory over the Riemann surface. For example, for smooth Riemann surfaces, preserving four-dimensional $\mathcal N=2$, the anomaly coefficients take the form\footnote{See \cite{Bah:2011vv,Bah:2011je} for a $\mathcal N=1$ generalization of this construction.}
	\begin{equation}
	\begin{aligned}
		a_{\mathcal N=2} = \frac{g-1}{12} (4N^3 -\frac32 N - \frac52)\,,\qquad c_{\mathcal N=2} = \frac{g-1}{12} (4N^3 - 2N - 2)\,.
	\end{aligned}
	\end{equation}
	Once again we observe the same cubic structure as we did in the previous section, without any quadratic term or $1/N$ corrections. We note that when the Riemann surfaces are punctured, there are additional $N^2$ terms in the anomaly coefficients \cite{Gaiotto:2009gz}, similar to the six-dimensional $D_N$ series. In this case, however, it is not clear that the $N^2$ terms should be removed through a redefinition of the flux quantisation. Instead, the punctured surfaces introduce new degrees of freedom that genuinely contribute at order $N^2$. These new degrees of freedom can be seen as open strings in the type IIA limit (KK-monopoles in M-theory) that provide Yang-Mills interactions at low energies. 

	\subsection{Five-dimensional SYM and giant gravitons}
	Another way to study the (2,0) theory is by compactifying the theory on a circle, which in the infrared reduces the theory to five-dimensional maximal SYM with gauge group $\U(N)$. It was argued that the sphere partition function of the latter should be identified with the Schur-like index of the $(2,0)$ theory, i.e. the partition function on $S^5 \times S^1_\beta$ \cite{Kim:2012ava,Kim:2012qf}
\be
Z(N,\beta)  = q^{E_c}{\cal I}_N(q)\,,\qquad {\cal I}_N(q) = \prod_{n=1}^N\prod_{m=0}^\infty \f{1}{1-q^{m+n}}\,,
\ee
and $E_c = N^3/6-N/8$ is the Casimir energy. It is convenient to separate out the $N\to\infty$ limit of the index and write
\be
Z(N,\beta)= q^{E_c}{\cal I}_\infty \e^{-F^{\text{np}}(N,\beta)}\,,
\ee
where $F^{\text{np}}$ can be viewed as the non-perturbative correction to the free energy while the perturbative part is given by $F^{\text{p}}(N,\beta) = \beta E_c - \log {\cal I}_\infty$ and is indeed \emph{perturbative} in $N$. The non-perturbative free energy admits the convenient expansion 
\be\label{Eq: Fnp 5d}
F^{\text{np}}(N,\beta)=-\sum_{n,m=0}^\infty \log(1-q^{1+N+m+n}) = \sum_{n=1}^\infty \f{\e^{-nN\beta}}{4n \sinh^2(n\beta/2)}\,.
\ee
This expression can be exponentiated and expanded to reproduce the giant graviton expansion in \cite{Arai:2020uwd}.
Note that, similar to the holographic equality in \eqref{Eq: Z equals J}, holography dictates 
\begin{equation}
	\mathcal Z_{\text{M2}}(N,\beta) = F(N,\beta)\,,
\end{equation}
and consequently we find that the series $F^{\text{np}}(N,\beta)$ should be captured by non-degenerate M2-branes wrapping $S^1_\beta$ and an $S^2\subset S^4$ $n$ times.
	The first term was reproduced on the gravitational side through a one-loop quantisation of a non-degenerate M2-brane \cite{Gautason:2023igo,Beccaria:2023sph} wrapping the compact three-cycle once. We conclude from \eqref{Eq: Fnp 5d}, as we did in the previous section, that the contribution of a quantised multi-wrapped M2-brane can be re-expressed as a rather simple transformation of the single wrapped brane, namely:
	\begin{equation}
		Z_{\text{M2}}^{(n)}(N,\beta) = n^{1-\chi} Z_{\text{M2}}^{(1)}(N,n\beta)\,,
	\end{equation}
	with once again $\chi = 2$ denoting the Euler characteristic of the two-dimensional base-space over which the M-theory circle is fibred. 
	Even though subleading saddles have not been fully explored (see however \cite{Gautason:2023igo} for a discussion of the $\beta\to0$ limit and the dual type IIA string), the QFT answer does suggest they should also possess one-loop exact partition functions as there are no further $1/N$ corrections. We remind the reader that when we reviewed non-degenerate M2-branes in AdS$_4\times S^7$ we also argued that they are one-loop exact in the M2-ensemble. 
	
	Finally, supersymmetric localisation of five-dimensional SYM also allows to compute the expectation value of Wilson line operators. Uplifting this result to six-dimensions translates the answer to the expectation values of defect operators wrapping $S^1_\beta \times S^1 \subset S^1_\beta \times S^5$, whose expectation value is \cite{Kim:2012ava,Kim:2012qf}
	\begin{equation}\label{Eq: WL in 2,0}
		\langle W \rangle = \frac{1}{2\sinh \beta/2 }\rme^{\beta N} - \frac{1}{2\sinh \beta/2} \,.
	\end{equation}
	In the bulk these operators correspond to M2-branes wrapping an AdS$_{3} \subset \text{AdS}_7$, with $S^1_\beta \subset \text{AdS}_3$ \cite{Gautason:2021vfc,Beccaria:2023sph}. And recently it was shown that a one-loop quantisation of these branes fully reproduces the leading saddle in \eqref{Eq: WL in 2,0}, making its partition function once again one-loop exact. The subleading saddle has zero classical action, but should have a one-loop exact partition function.
	
	We found in all previous examples that the supersymmetric partition function of a multi-wound M2-brane takes a particularly simple form in terms of a single-wound M2-brane \eqref{Eq: multiwound vs single wound}. Extrapolating this result to the defect operator expectation value discussed here we find that
	\begin{equation}\label{Eq: multi wound WL in 2,0}
		\langle W_n \rangle = \frac{1}{2\sinh n\beta/2 }\rme^{n\beta N} - \frac{1}{2\sinh n\beta/2} \,,
	\end{equation}
	which is a particularly interesting result to test via supersymmetric localisation in the field theory dual.

	\section{Final remarks}
	In this note we discussed the generalisation of the string generating functional, where the target space fields act as background sources \cite{Fradkin:1984pq,Fradkin:1985fq,Fradkin:1985ys}, to M2-branes. The existence of the M2-brane generating functional is a natural consequence of the type IIA/M-theory duality, but further evidence for its existence comes from the match of different holographic observables involving the semi-classical quantisation of M2-branes, see e.g. \cite{Giombi:2023vzu,Beccaria:2023sph,Beccaria:2023ujc,Gautason:2025per}. We emphasised that the M2-brane partition function depends only on the metric and three-form $A_3$. The conjugate variable $\star G_4$, or the six-form potential $A_6$, however, is not a parameter and is therefore unfixed. We thus denoted the partition function as being in the M2-ensemble whereas in the conjugate ensemble, which we denote as the M5-ensemble, the six-form $A_6$ is fixed. 

	In the setups we have studied there is a direct relation between M-theory and type IIA string theory, and consequently it seems that the quantisation of M2-branes is under better control (compared to the the quantisation of M5-branes), making the M2-ensemble a more natural choice in which one computes quantum gravitational observables. This implies for example that in ABJM holography (and other related examples) we naturally compute observables in the grand canonical ensemble where $N$, the rank of the gauge group, is unfixed. An evidence for this can already be seen at leading order where the direct evaluation of the elven-dimensional on-shell action does \emph{not} agree with the QFT prediction in the canonical ensemble but rather with its grand canonical partition function (see section \ref{sec:ensembles_in_ads_4_holography}). To give further evidence we could evaluate the leading perturbative correction to the eleven-dimensional action, which consists of two superinvariants at eight derivative order. Unfortunately these terms have not been fully supersymmetrized and hence the full eight derivative term is still unknown. We are therefore currently unable to verify this aspect of our proposal. 
	An alternative approach would be to directly quantize degenerate M2-branes whose leading one-loop contribution would reproduce the two-derivative supergravity answer but higher loop corrections essentially access these higher derivative terms.

Through our study of the M2-brane generating functionals in various smooth holographic backgrounds of different dimensions, we found that it generally takes a remarkably simple form, namely
	\begin{equation}\label{Eq: cubics}
		\mathcal Z_{\text{M2}}(\mu,\mathbf{q}) = a(\mathbf{q}) \mu^3 + b(\mathbf{q}) \mu + c(\mathbf{q}) + {\cal O}(\e^{-\mu})\,,
	\end{equation}
	where $\mathbf{q}$ denotes various parameters that the background metric depends on and $\mu$  determines the three-form $A_3$ and is generally related to the planck length via $\mu\sim1/\ell_p^3$. The evidence for the cubic structure primarily comes from exact computations in supersymmetric quantum field theories that admit a holographic description in M-theory. Directly in M-theory we can argue for the first three terms based on derivative counting and what is known about the structure of the string- and M-theory effective action, however explicitly computing the $b$ and $c$ coefficients is currently out of reach as alluded to above. It is very surprising that in the cases we considered there are no further any $1/\mu$ corrections. Without some sort of non-renormalization theorems, proving this fact directly in M-theory requires similar non-perturbative methods as to those that are used on the QFT side of the duality, which are currently poorly understood. 
	
	Assuming that the partition function truncates to a cubic polynomial as above has rather interesting consequences for AdS$_4$ holography, namely it implies that while the grand canonical partition function of the dual field theory is a simple cubic, in the canonical ensemble the partition function is given by an Airy function (plus exponentially suppressed terms). Furthermore, the argument of the Airy function is $N$ shifted by the coefficient $b(\mathbf{q})$. This follows directly from the Laplace transform between the canonical and grand canonical ensembles. Our message is that we should not attempt to derive the Airy function from supergravity (and its corrections). Rather, we should only attempt to derive the cubic polynomial.
	Taking instead the AdS$_7 \times S^4$ background, it is known that anomalies and partition functions of the dual six-dimensional $(2,0)$ theory take a similar cubic form. 
	This shows that the M2-brane partition function in AdS$_7$ geometries takes the same form as in AdS$_4$ geometries upon a correct identification of the parameter $\mu$, which in the latter case is simply the dual field theory parameter $N$. 
	The fact that no ensemble switch is needed is a consequence of the fact that the rank of the gauge group $N$ is here dual to the $G_4$ flux and hence the M2-brane partition function directly depends on it. 
	
	Another interesting observation that we found is that the partition function of supersymmetric non-degenerate M2-branes in the holographic backgrounds studied is one-loop exact. This observation was discussed explicitly for M2-branes that are dual to instanton corrections in both three-dimensional and six-dimensional CFTs, and also for Wilson lines and defect operators in the respective examples.
	This one-loop exactness has been verified through the semi-classical quantisation of a single M2-branes in the corresponding setups in \cite{Gautason:2023igo,Beccaria:2023ujc,Beccaria:2023sph,Gautason:2025per}. Furthermore, we observed that multi-wound M2-branes have a partition function that can be written in terms of the single-wound M2-brane partition function, namely:
	\begin{equation}
		\mathcal Z_{\text{M2}}^{(n)}(\mu,\beta) = n^{1-\chi} \mathcal Z_{\text{M2}}^{(1)}(\mu,n \beta)\,,
	\end{equation}
	where $n$ is the winding number, $\mu$ is the parameter fixing the background source $A_3$, $\beta$ is the periodicity of the M-theory circle that the brane wraps, and finally $\chi$ is the Euler characteristic of the base space of the M2-brane over which the M-theory circle is fibred. These facts could potentially be explained by the same argument that explains the cubic structure of the perturbative partition function but we leave that to important future research which surely would lead to significant advancement of our understanding of top-down holography.

\section*{Acknowledgements}
We are thankful to Stefan Kuryland, Alexia Nix, and Thomas Van Riet for useful discussions, and Nikolay Bobev and Arkady Tseytlin for useful comments on this preprint. FFG is supported in part by the Icelandic Research Fund under grant 228952-053. JvM is supported by the STFC Consolidated Grant ST/X000575/1. JvM is grateful for the continuous hospitality of the ITF at the KU Leuven.

\bibliography{LocalisingM2sbib.bib}
\bibliographystyle{JHEP}
	
\end{document}